\documentclass{article}
\usepackage{graphicx} 
\usepackage{amsmath}
\usepackage{amssymb}
\usepackage{mathtools}
\usepackage{amsfonts}
\usepackage[space]{cite}
\usepackage[english]{babel}

\newtheorem{claim}{Claim}

\usepackage{xcolor}

\newcommand{\re}{{\rm e}}
\newcommand{\id}{\mathbb{I}}
\newcommand{\pha}{\alpha}

\usepackage{tikz}
\usetikzlibrary{quantikz}
\usepackage{adjustbox}
\usepackage{caption}
\usepackage{subcaption}

\usepackage{ulem}

\usepackage{MnSymbol}

\makeatletter
\newcommand*\dashline{\rotatebox[origin=c]{90}{$\dabar@\dabar@\dabar@$}}
\makeatother

\newcommand{\RN}[1]{%
  \textup{\uppercase\expandafter{\romannumeral#1}}%
}

\numberwithin{equation}{section}

\begin{document}
\begin{titlepage}

\begin{center}

{\LARGE $q$-analog qudit Dicke states}\\
\vspace{1in}

\large David Raveh\footnote{\tt dxr921@miami.edu}\quad and \quad Rafael I. Nepomechie\footnote{\tt nepomechie@miami.edu}\\[0.2in] 
Physics Department, PO Box 248046\\[0.2in] 
University of Miami, Coral Gables, FL 33124 USA
\end{center}

\vspace{.5in}

\begin{abstract}

Dicke states are completely symmetric states of multiple
qubits (2-level systems), and qudit Dicke states are their $d$-level generalization. 
We define here $q$-deformed qudit Dicke states using the quantum algebra $su_q(d)$. We show that these states can be compactly expressed as a weighted sum over permutations with $q$-factors involving the so-called inversion number,  an important permutation statistic in Combinatorics. We use this result to compute the bipartite entanglement entropy of these states. We also discuss the preparation of these states on a quantum computer, and show that introducing a $q$-dependence does not change the circuit gate count.
\end{abstract}

\end{titlepage}

\setcounter{footnote}{0}

\section{Introduction}\label{sec:intro}

Completely symmetric multi-qubit states are called
Dicke states. These highly entangled states
have long been exploited for 
a wide variety of tasks in quantum information and computation, including quantum networking, quantum metrology, quantum compression, quantum tomography, and quantum optimization see e.g. \cite{Dicke:1954zz, Murao:1999, 
Ozdemir:2007,
Prevedel:2009,
Toth:2010, Toth:2012, 
Farhi:2014, Ouyang:2014, Ouyang:2021}, 
with specific interest in their entanglement characterizations \cite{Popkov:2004, Latorre:2004qn, 
Munizzi:2023ihc}, as well as methods of constructing these states in the lab
\cite{Kiesel:2007, Prevedel:2009,  Wieczorek:2009} and on a quantum computer 
\cite{Bartschi2019, Mukherjee:2020, 
Aktar:2021, Bartschi:2022}.
There has been growing interest in using $d$-level systems called qudits for quantum computation, see e.g. \cite{Wang:2020, Goss:2022, Hrmo:2023, Morvan:2021, Ringbauer:2022, Roy:2022} and references therein. Completely symmetric multi-qudit states, which we call qudit Dicke states (also known as generalized Dicke states, or symmetric basis states), have also been studied for many years, see e.g \cite{Wei:2003, Popkov:2005,
Hayashi:2008, Wei:2008, Zhu:2010, Carrasco:2015sxh, Li:2021, Nepomechie:2023lge}.

In order to specify such multi-qudit Dicke states, it is convenient to introduce some notation.
Let $\vec k=(k_0,k_1,\dots, k_{d-1})$, where $k_i$ are non-negative integers, and $n=\sum_{i=0}^{d-1} k_i$ is the number of qudits. Let $M(\vec k)$ denote the multiset
\begin{equation}
M(\vec k)	=\{ \underbrace{0, \ldots, 0}_{k_{0}}, \underbrace{1, 
\ldots, 1}_{k_{1}}, \ldots, \underbrace{d-1, \ldots, d-1}_{k_{d-1}}\},
\label{multiset}
\end{equation}
where $k_i$ is the multiplicity of $i$, and let $\mathfrak{S}_{M(\vec k)}$ denote the set of permutations of $M(\vec k)$, which has cardinality equal to the multinomial \cite{Stanley2011}
\begin{equation}
{n \choose \vec k}={n \choose k_0,k_1,\dots,k_{d-1}}=\frac{n!}{\prod_{i=0}^{d-1}k_i!} \,.
\label{multinomial}
\end{equation}
The qudit Dicke states can then be expressed as \cite{Wei:2003, Popkov:2005, Hayashi:2008, Wei:2008, Zhu:2010, Carrasco:2015sxh, Li:2021, Nepomechie:2023lge}
\begin{equation}
|D^n(\vec k)\rangle=
\frac{1}{\sqrt{{n \choose \vec k}}}
\sum_{w\in \mathfrak{S}_{M(\vec k)}}
|w\rangle \,,
\label{Dicke}
\end{equation}
where we sum over all permutations $w$ of $M(\vec k)$.
For example, the qudit Dicke state with $d=3$, $n=4$ and $\vec k=(1,2,1)$ is 
\begin{equation}
\begin{split}
|D^4(1,2,1)\rangle&=\frac{1}{\sqrt{12}}
\Big(|0112\rangle+|1012\rangle
+|0121\rangle+|1102\rangle
+|0211\rangle+|1021\rangle\\&
+|1120\rangle+|1201\rangle
+|2011\rangle+|1210\rangle
+|2101\rangle+|2110\rangle
\Big) \,,
\label{DickeExample}
\end{split}
\end{equation}
where as usual the tensor product is understood, e.g. $|0112\rangle = |0\rangle \otimes |1\rangle \otimes |1\rangle \otimes |2\rangle$, and the computational basis states are denoted by $|0\rangle\,, |1\rangle\,, \dots\,, |d-1\rangle$, see \eqref{basis}.

We consider in this paper a $q$-analog of qudit Dicke states, which we call $q$-qudit Dicke states. 
In principle, there are many possible one-parameter ($q$) deformations of qudit Dicke states, which reduce to
\eqref{Dicke} in the limit $q \to 1$. We consider here a particular deformation based on Quantum Groups, which has a number of attractive features, as we will see. The qubit case ($d=2$) has been considered in \cite{Li:2015}. Our main result is a formula \eqref{Dickesum} that generalizes \eqref{Dicke} for $q$-qudit Dicke states, namely
\begin{equation}
|D_q^n(\vec k)\rangle=
\frac{1}{\sqrt{{n \brack \vec k}}}
\sum_{w\in \mathfrak{S}_{M(\vec k)}}
q^{J(\vec k)/2-\text{inv}(w)} |w\rangle \,,
\label{mainresult}
\end{equation}
where $\text{inv}(w)$ denotes the inversion number and $J(\vec k)$ the maximum inversion number, see Sections \ref{sec:qgroup} and \ref{sec:qDicke} for more details.
We use this result to compute the entanglement entropy of $q$-qudit Dicke states \eqref{EE}.  
The inversion number is an important permutation statistic in Combinatorics \cite{Stanley2011}; it is interesting to see that it and the related $q$-combinatorial identities \eqref{qinvidentity} and \eqref{qVandermonde} appear in Quantum Information. 
We expect that $q$-qudit Dicke states will find  applications similar to those of their undeformed ($q=1$) counterparts, but with the advantage of having available a free parameter $q$ as an additional degree of freedom. 

This paper is organized as follows.  In Section \ref{sec:qgroup} we
briefly review $su_q(d)$, the $q$-deformation of the $su(d)$ algebra.
In Section \ref{sec:qDicke} we use the generators of $su_q(d)$ to
define $q$-qudit Dicke states, and we note their key properties.  In
particular, we note a recursion relation for $q$-qudit Dicke states
\eqref{Dickerecursion} (proved in Appendix \ref{sec:recursionproof}),
which we use to derive the result \eqref{mainresult}.  In Section
\ref{sec:EE} we compute the bipartite entanglement entropy of
$q$-qudit Dicke states, using the Schmidt decomposition
\eqref{schmidt} that is proved with the help of \eqref{mainresult} in
Appendix \ref{sec:schmidtproof}.  In Section \ref{sec:algorthim} we
take advantage of the recursive nature of the $q$-qudit Dicke states
to formulate an efficient deterministic method of preparing these
states on a quantum computer.  In Section \ref{sec:conclusion} we
briefly discuss our results, and note some possible directions for
further investigation.  Mathematica and Qiskit codes supporting these 
findings are provided as Supplementary Material.

\section{Review of $su_q(d)$}\label{sec:qgroup}

We briefly review here the $su_q(d)$ algebra with $q>0$, following \cite{Jimbo:1985vd, Chari:1994pz}, which we will use in the following section to define $q$-qudit Dicke states.

For a single qudit, the generators of the quantum algebra $su_q(d) = U_q(su(d))$
are the same as for the classical (undeformed) algebra. In particular, we define the Cartan generators  
\begin{equation}
H_i^{(1)}=e_{i,i}-e_{i+1,i+1}, \qquad i=1, 2,\dots, d-1\,,
\label{Cartan1}
\end{equation}
and the Chevalley generators \begin{equation}
X^{+(1)}_i =e_{i,i+1}, \qquad
X^{-(1)}_i =e_{i+1,i}, \qquad i=1, 2, \dots, d-1 \,,
\label{Chevalley1}
\end{equation}
where $e_{ij}$ are elementary $d\times d$ matrices with matrix elements $\left(e_{ij}\right)_{ab}=\delta_{ai}\delta_{bj}$.
Other generators can be obtained by taking commutators.

For $n$ qudits, living in the vector space $\left(\mathbb{C}^d\right)^{\otimes n}$,
the corresponding generators are given by the coproducts \cite{Jimbo:1985vd, Chari:1994pz}
\begin{align}
H_i^{(n)} &=\sum_{j=0}^{n-1} \,
\underset{\mathclap{\substack{\uparrow \\ 0}}}{\id}
\otimes\dots\otimes 
\id\otimes
\underset{\mathclap{\substack{\uparrow \\ j}}}{H_i^{(1)}}
\otimes \id\otimes\dots\otimes 
\underset{\mathclap{\substack{\uparrow \\ n-1}}}{\id} \,, \label{Cartann}\\
X^{\pm(n)}_i &=\sum_{j=0}^{n-1} 
\underset{\mathclap{\substack{\uparrow \\ 0}}}{q^{H^{(1)}_i/2}}
\otimes\dots\otimes 
q^{H^{(1)}_i/2}\otimes
\underset{\mathclap{\substack{\uparrow \\ j}}}{X^{\pm(1)}_i}
\otimes q^{-H^{(1)}_i/2}\otimes
\dots\otimes 
\underset{\mathclap{\substack{\uparrow \\ n-1}}}{q^{-H^{(1)}_i/2}} \,,
\label{Chevalleyn}
\end{align}
$i=1, 2, ..., d-1$, 
where $\id$ is the $d \times d$ identity matrix,
$H_i^{(1)}$ and $X^{\pm(1)}_i$ occur at the $j$th location, and we sum $j$ over all $n$ positions. It is important to note for the next section
that $X^{\pm(n)}_i$ may be defined recursively
\begin{equation}
X^{\pm(n)}_i=
\left(q^{H^{(1)}_i/2}\right)^{\otimes (n-1)}\otimes X^{\pm(1)}_i+
X^{\pm(n-1)}_i\otimes q^{-H^{(1)}_i/2} \,.
\label{gensrecursive}
\end{equation}
The algebra includes the commutator relations
\begin{align}
\left[H^{(n)}_i,X^{\pm(n)}_j \right]&=
\pm(2\delta_{i,j}-\delta_{i-1,j}
-\delta_{i+1,j})X^{\pm(n)}_j \,,
\\
\left[X^{+(n)}_i,X^{-(n)}_j \right]&=
\delta_{ij}\left[H^{(n)}_i\right],
\end{align}
where we use the bracket notation
\begin{equation}
[x]=\frac{q^x-q^{-x}}{q-q^{-1}}\,,
\label{bracket}
\end{equation}
so that $[x]$ is symmetric under $q\to1/q$. It is clear that $[x]\to x$ as $q\to1$, so that $[x]$ is a $q$-deformation of $x$. 
For more details about $su_q(d)$, see 
e.g. \cite{Jimbo:1985vd, Chari:1994pz}.

\section{$q$-qudit Dicke states}\label{sec:qDicke}

We define here $q$-qudit Dicke states, and note some of their key
properties, including the recursion formula \eqref{Dickerecursion},
the sum formula \eqref{Dickesum} and its extension to complex values
of $\mathbf{q}$ \eqref{Dickesum2}, and the duality symmetry
\eqref{Dickesymmetry}.

The single-qudit computational basis states are given as usual by the $d$-dimensional vectors
\begin{equation}
	|0\rangle = \begin{pmatrix}1\\0\\ \vdots\\0\end{pmatrix} \,, \qquad
	|1\rangle = \begin{pmatrix}0\\1\\ \vdots\\0\end{pmatrix} \,, \ldots \,, \qquad
	|d-1\rangle = \begin{pmatrix}0\\0\\ \vdots\\1\end{pmatrix} \,.
 \label{basis}
\end{equation}
The single-qudit operators defined in \eqref{Cartan1}, \eqref{Chevalley1} perform the following mappings on these basis states:
\begin{align}
H^{(1)}_i|j\rangle &=
\delta_{i,j+1}|j\rangle-\delta_{i,j}|j\rangle \,,
\label{Hmap} \\
X^{- (1)}_i |j\rangle &=\delta_{i,j+ 1}|j + 1\rangle\,, \qquad 
X^{+ (1)}_i |j\rangle =\delta_{i,j}|j - 1\rangle\,.
\label{Xmap} 
\end{align}

\subsection{Definition and recursive property}

We use the $su_q(d)$ generators \eqref{Cartann} and \eqref{Chevalleyn} to define the $q$-qudit Dicke states for fixed $q>0$ and $\vec k$ by the ``operator formula'' \footnote{We occasionally place a subscript $n$ on a ket, such as $|0\dots0\rangle_n$ in \eqref{DickeV1}, to clarify that it is a state of $n$ qudits.}
\begin{align}
|D_q^n(\vec k)\rangle&=
\frac{1}{\sqrt{{n \brack \vec k}}}
\frac{1}{[k_1]!\dots[k_{d-1}]!}
{X^{-(n)}_1}^{k_1}
{\left[X^{-(n)}_2, X^{-(n)}_1\right]}^{k_2}
{\left[X^{-(n)}_3, \left[X^{-(n)}_2,X^{-(n)}_1 \right] \right]}^{k_3} \nonumber\\
&\times \dots{\left[X^{-(n)}_{d-1},\dots \left[X^{-(n)}_2,X^{-(n)}_1 \right]\dots \right]}^{k_{d-1}}
|0\dots0\rangle_n \,,
\label{DickeV1}
\end{align}
where ${n \brack \vec k}$ denotes the $q$-multinomial
\begin{equation}
{n \brack \vec k}={n \brack k_0,k_1,\dots,k_{d-1}}=\frac{[n]!}{\prod_{i=0}^{d-1}[k_i]!} \,,
\label{qmultinomial}
\end{equation}
and the $q$-factorial is defined for
non-negative integers $n$ by
\begin{equation}
[n]!=[n][n-1]\dots[1],\quad\text{with}\quad[0]!=1 \,.
\end{equation}
Indeed, for a single qudit, it is easy to see that
the nested commutator $\left[X^{-(1)}_j,\dots \left[X^{-(1)}_2,X^{-(1)}_1 \right]\dots\right]$ lowers the basis state $|0\rangle$ to $|j\rangle$, i.e.
\begin{align}
    X^{-(1)}_1\, |0\rangle &= |1\rangle \,, \nonumber\\
    \left[X^{-(1)}_2, X^{-(1)}_1\right]\, |0\rangle &= |2\rangle \,, \nonumber \\
    & \vdots \nonumber \\
\left[X^{-(1)}_{d-1},\dots \left[X^{-(1)}_2,X^{-(1)}_1 \right]\dots\right] |0\rangle &= |d-1\rangle   \,,
\label{Dickeinitial}
\end{align}
where the basis states are defined in \eqref{basis}.
Hence, for $q\rightarrow 1$, the formula \eqref{DickeV1} can be seen to give the qudit Dicke state \eqref{Dicke}.\footnote{A proof of this fact follows from the $q\rightarrow 1$ limit of \eqref{Dickesum}.}

The $q$-qudit Dicke state \eqref{DickeV1} satisfies the important recursion
\begin{equation}
|D_q^n(\vec k)\rangle
=\sum_{s=0}^{d-1}
\sqrt{\frac{[k_s]}{[n]}}
q^{\frac{1}{2}\left(\sum_{r=0}^{s-1}k_r-\sum_{r=s+1}^{d-1}k_r \right)} |D_q^{n-1}(\vec k-\hat s)\rangle\otimes|s\rangle \,,
\label{Dickerecursion}
\end{equation}
with the initial conditions $|D_q^1(\hat s)\rangle=|s\rangle$ as given in \eqref{Dickeinitial},
where $\hat s$ is the $s$th unit vector in $d$ dimensions 
\begin{equation}
    \hat s = (\underset{\mathclap{\substack{\uparrow \\ 0}}}{0}, 
    \ldots, 0, \underset{\mathclap{\substack{\uparrow \\ s}}}{1}, 
    0, \ldots, \underset{\mathclap{\substack{\uparrow \\ d-1}}}{0})
    \quad\text{so that} \quad
    \vec k-\hat s=(k_0,\dots,k_s-1,\dots,k_{d-1}) \,.
    \label{shat}
\end{equation}
In the sum over $s$ in \eqref{Dickerecursion},
we implicitly skip over $s=i$ if $k_i=0$, since  $|D_q^{n-1}(\vec k-\hat i)\rangle$ is then not defined. A detailed proof of this recursion is given in Appendix \ref{sec:recursionproof}; the main idea is to exploit the recursive nature of $X^{-(n)}_i$ in \eqref{gensrecursive}. As we shall see in Section \ref{sec:algorthim}, it is due to the recursive nature of the $q$-qudit Dicke states that these states can be efficiently generated on a quantum computer.

\subsection{Sum formula}

The $q$-qudit Dicke states can be written explicitly in terms of permutations, generalizing the $q=1$ result \eqref{Dicke}. To do so, we need to borrow a notion from Combinatorics (see, e.g. \cite{Stanley2011}). 
Every permutation $w$ of $M(\vec k)$ can be regarded as a ``word'' with $n$ ``letters'' $w_i$ in the alphabet $\{0,1,\ldots,d-1\}$, i.e. $w=w_1 w_2\dots w_n$. We refer to the permutation in weakly increasing order, satisfying $w_i\leq w_j$ for all $i<j$, e.g. $0112$, as the {\it identity permutation} $\re(\vec{k})$. The {\it inversion number} of a permutation $w$, written $\text{inv}(w)$, represents the minimum number of adjacent transpositions it takes to go from the identity permutation to $w$. The inversion number of $w$ can be efficiently computed by
\begin{equation}
\text{inv}(w)=\sum_{j=1}^n z_j \,,
\end{equation} 
where $z_j$ equals the number of letters $w_i>w_j$ satisfying $i<j$; that is $z_j$ is the number of letters to the left of $w_j$ larger than $w_j$.
For example, for $w=1201$, we have $w_1=w_4=1,\,w_2=2,\,w_3=0$; thus $z_1=z_2=0,\,z_3=2,\, z_4=1$; and hence $\text{inv}(w)=3$.

We let $J(\vec k)$ denote the maximum inversion number of all permutations of $M(\vec k)$; clearly this corresponds to the permutation in reverse-order of the identity permutation $\re(\vec k)$ (i.e. weakly decreasing). Thus 
\begin{equation}
J(\vec k)=\max_{w\in \mathfrak{S}_{M(\vec k)}}\text{inv}(w)=\sum_{0\leq i<j\leq d-1}k_i k_j\,,
\label{Jresult}
\end{equation}
where the term $k_i k_j$ with $i<j$ represents the $k_j\, j$'s to the left of the $k_i\, i$'s in the reverse of the identity. 

We claim that the $q$-qudit Dicke states \eqref{DickeV1} may be expressed by the ``sum formula'' 
\begin{equation}
|D_q^n(\vec k)\rangle=
\frac{1}{\sqrt{{n \brack \vec k}}}
\sum_{w\in \mathfrak{S}_{M(\vec k)}}
q^{J(\vec k)/2-\text{inv}(w)} |w\rangle \,,
\label{Dickesum}
\end{equation}
where we sum over all permutations $w$ of $M(\vec k)$. 
For example, the $q$-qudit Dicke state corresponding to \eqref{DickeExample} with $\vec k=(1,2,1)$ is 
\begin{align}
&|D_q^4(1,2,1)\rangle=\frac{1}{\sqrt{q^5+2q^3+3q+3q^{-1}+2q^{-3}+q^{-5}}}
\Big(q^{5/2}|0112\rangle
+q^{3/2}|1012\rangle
+q^{3/2}|0121\rangle\nonumber\\
&\quad +q^{1/2}|1102\rangle
+q^{1/2}|0211\rangle
+q^{1/2}|1021\rangle
+q^{-1/2}|1120\rangle
+q^{-1/2}|1201\rangle
+q^{-1/2}|2011\rangle\nonumber\\
&\quad +q^{-3/2}|1210\rangle
+q^{-3/2}|2101\rangle
+q^{-5/2}|2110\rangle
\Big) \,. 
\label{DickeExampleq}
\end{align}
To prove that the sum formula \eqref{Dickesum} is valid, all we need to do is show that this formula satisfies the recursion in \eqref{Dickerecursion} along with the same initial conditions. Clearly, \eqref{Dickesum} has the same initial conditions $|D_q^1(\hat s)\rangle=|s\rangle$. Note that any permutation $w$ that ends with the letter $s$ can be expressed as $|w\rangle_n=|w_s\rangle_{n-1}\otimes|s\rangle$, where $w_s$ is a permutation of $M(\vec k - \hat s)$. The inversion numbers for $w$ and $w_s$ are related by \begin{equation}
\text{inv}(w)=\text{inv}(w_s)+\sum_{r=s+1}^{d-1} k_r\,,
\end{equation}
where the sum denotes the number of adjacent transpositions it takes to move the right-most $s$ in the identity permutation $|\re(\vec k)\rangle_n$ to the $n$th position, obtaining $|\re(\vec k-\hat s)\rangle_{n-1}\otimes|s\rangle$. Let $J(\vec k-\hat s)$ denote the corresponding maximum inversion number over all $w_s$. Then $J(\vec k)=J(\vec k-\hat s)+\sum_{r\neq s} k_r$ because 
\begin{equation}
J(\vec k)=\sum_{\substack{i<j\\i\neq s\\j\neq s}}k_i k_j+\sum_{r\neq s}k_r k_s
=\sum_{\substack{i<j\\i\neq s\\j\neq s}}k_i k_j+\sum_{r\neq s}k_r(k_s-1)+\sum_{r\neq s}k_r=J(\vec k-\hat s)+\sum_{r\neq s}k_r.
\end{equation}
The sum $\sum_{r\neq s}k_r$ represents the number of non-$s$ letters that cross over the right-most $s$ in $\re(\vec k)$ when reversing the identity permutation $\re(\vec k)$.
Substituting $J(\vec k)$, $\text{inv}(w)$, and $|w\rangle$ in terms of $J(\vec k-\hat s)$, $\text{inv}(w_s)$, and $|w_s\rangle\otimes|s\rangle$ in \eqref{Dickesum}, using the fact ${n \brack \vec k} = \frac{[n]}{[k_s]} {n-1 \brack \vec k - \hat{s}}$,
and summing over all possible $s$, we see that \eqref{Dickesum} satisfies recursion \eqref{Dickerecursion}.
That the factor with $\text{inv}(w)$ is present in front of $|w\rangle$ in \eqref{Dickesum}
is not entirely surprising, as the proof of the recursion of the $q$-qudit Dicke states (see Appendix \ref{sec:recursionproof}) makes use of counting interchanges of operators that each produces some $q$-factor. 

The sum formula \eqref{Dickesum} makes clear the  orthogonality of any pair of $q$-qudit Dicke states with different $\vec k$'s. The identity\footnote{\label{inv}This identity has a long and interesting history recounted in \cite{Stanley2011}. 
In Eq. (1.68) of \cite{Stanley2011}, this identity is written in the form 
\begin{equation}
\sum_{w\in \mathfrak{S}_{M(\vec k)}}
\Tilde q^{\text{inv}(w)}=\boldsymbol{{n \choose \vec k}}\,,
\label{mathid}
\end{equation}
where the boldface denotes the $\Tilde q$-multinomial using a definition of $\Tilde q$-deformation that is 
different from \eqref{bracket}, namely
$\boldsymbol{(x)}=\frac{1-\Tilde q^x}{1-\Tilde q}= \frac{\Tilde{q}^{x/2}}{\Tilde{q}^{1/2}}[x]_{q=\Tilde{q}^{-1/2}}$ where the $q$ in $[x]$ is evaluated at $\Tilde{q}^{-1/2}$. Computing $\boldsymbol{(n)!}$ using $\sum_{i=1}^n i=n(n+1)/2$ and using $2J(\vec k)=n(n+1)-\sum_{i=0}^{d-1}k_i(k_i+1)$ gives $\boldsymbol{{n \choose \vec k}}=\Tilde{q}^{J(\vec k)/2}{n \brack \vec k}_{q=\Tilde{q}^{-1/2}}$. Substituting this into \eqref{mathid} and letting 
$\Tilde{q}=q^{-2}$ leads to our form \eqref{qinvidentity}.
}
\begin{equation}
\sum_{w\in \mathfrak{S}_{M(\vec k)}}
q^{J(\vec k)-2\, \text{inv}(w)}={n \brack \vec k}  
\label{qinvidentity}
\end{equation}
confirms that the $q$-qudit Dicke states \eqref{Dickesum} are normalized to unity, and thus are orthonormal. This identity also shows that the $q$-multinomial can be expressed as a Laurent polynomial in $q$ with non-negative integer coefficients cf. the denominator in \eqref{DickeExampleq}, which is a nontrivial generalization of the fact 
$[n]=q^{n-1}+q^{n-3}+\cdots+q^{3-n}+q^{1-n}$ for positive integers $n$. 

\subsection{Complex $\mathbf{q}$}

We have so far assumed that the deformation parameter $q$ is positive $q>0$. Let us now briefly consider a complex deformation parameter $\mathbf{q} = q e^{i \pha}$, with $q = |\mathbf{q}|>0$ and $\pha$ real. 
It is not obvious how to define corresponding $\mathbf{q}$-qudit Dicke states; indeed, replacing $q$ by $\mathbf{q}$ in \eqref{DickeV1} could lead to complications if $\mathbf{q}$ is a root of unity,
i.e. $\mathbf{q}^p =1$ for some positive integer $p$, since then $[p]=0$.
We therefore define $\mathbf{q}$-qudit Dicke states instead by the recursion \eqref{Dickerecursion} with $q$ replaced by $\mathbf{q}$, i.e.
\begin{equation}
|D_\mathbf{q}^n(\vec k)\rangle
=\sum_{s=0}^{d-1}
\sqrt{\frac{[k_s]}{[n]}}
\mathbf{q}^{\frac{1}{2}\left(\sum_{r=0}^{s-1}k_r-\sum_{r=s+1}^{d-1}k_r \right)}|D_\mathbf{q}^{n-1}(\vec k-\hat s)\rangle\otimes|s\rangle \,,
\label{Dickerecursiongen}
\end{equation}
where the brackets are given by \eqref{bracket} with $q = |\mathbf{q}|>0$,
which is well-defined (after choosing the square-root branch cut)
even if $\mathbf{q}$ is a root of unity. It follows that 
\begin{equation}
|D_\mathbf{q}^n(\vec k)\rangle=
\frac{1}{\sqrt{{n \brack \vec k}}}
\sum_{w\in \mathfrak{S}_{M(\vec k)}}
\mathbf{q}^{J(\vec k)/2-\text{inv}(w)} |w\rangle \,,
\label{Dickesum2}
\end{equation}
where again the brackets are given by \eqref{bracket} with $q = 
|\mathbf{q}|>0$. 

As an example, let us consider the case $d=2\,, 
n=3\,, \vec{k}=(2,1)$ with $\mathbf{q}=e^{2\pi i/3}$. The brackets are to be computed with the absolute value of $\mathbf{q}$; 
that is, with $q=| \mathbf{q} | = 1$. Therefore, 
\begin{equation}
	{3 \brack 2\,, 1}_{q=| \mathbf{q}| =1}  = {3 \choose 2\,, 1}  = 3 \,, 
\end{equation}
and the $\mathbf{q}$-qudit Dicke state \eqref{Dickesum2} is given by
\begin{equation}
	|D^{3}_{\mathbf{q}}(2,1)\rangle = 
	\frac{1}{\sqrt{3}}\left( \mathbf{q}^{-1}|100\rangle + 
	|010\rangle + \mathbf{q}\, |001\rangle \right) \,.
\end{equation}	

Note that defining the $\mathbf{q}$-qudit Dicke states by 
\eqref{Dickerecursiongen} ensures that they are properly normalized
$\langle D_\mathbf{q}^n(\vec k)|D_\mathbf{q}^n(\vec k)\rangle=1$, where as usual $\langle D_\mathbf{q}^n(\vec k)|=\left(|D_\mathbf{q}^n(\vec k)\rangle\right)^\dagger$ and $\mathbf{q}^\dagger$ is given by the complex conjugate of $\mathbf{q}$.
The special case $\mathbf{q}=-1$, which corresponds to antisymmetric states, has been considered in the literature, see e.g. \cite{Wei:2003, Hayashi:2008, Li:2021, Bravyi:2003} and references therein.

\subsection{Duality symmetry}

The sum formula \eqref{Dickesum2} can be used to show 
that the $\mathbf{q}$-qudit Dicke states have
the ``duality'' symmetry 
\begin{equation}
|D^n_\mathbf{q}(\vec k)\rangle=
\mathcal{C}^{\otimes n}\, |D^n_{1/\mathbf{q}}({\rm rev}(\vec k))\rangle \,,
\label{Dickesymmetry}
\end{equation}
where 
\begin{equation}
{\rm rev}(\vec k)=(k_{d-1},k_{d-2},\dots,k_0)
\label{reverse}
\end{equation} 
denotes the reverse of $\vec k$, and $\mathcal{C}$ is the $d \times d$ antidiagonal ``charge conjugation'' matrix
\begin{equation}
    \mathcal{C} = \sum_{i=1}^d e_{i,d+1-i} 
    = \begin{pmatrix}
    & &  1 \\
    & \udots & \\
    1 & & 
    \end{pmatrix} \,, \qquad
\end{equation}
which performs the mapping 
\begin{equation}
   \mathcal{C}\, |j\rangle = |d-1-j\rangle \,, \qquad j = 0, 1, \ldots d-1 \,,
\end{equation}
on the single-qudit basis states \eqref{basis}. This is a generalization of the well-known duality property of qubit Dicke states $|D^n(n-l,l)\rangle = X^{\otimes n}\, |D^n(l,n-l)\rangle$.

\section{Entanglement entropy}\label{sec:EE}

Quantum entanglement is a key resource of Quantum Information. An important measure of quantum entanglement is the bipartite entanglement entropy, see e.g. the reviews \cite{Horodecki:2009zz, Guhne:2009}. For qubit Dicke states ($d=2\,, q=1$), the bipartite entanglement entropy was computed in \cite{Popkov:2004, Latorre:2004qn, Munizzi:2023ihc}; this result was generalized for qudit Dicke states (general $d$, but $q=1$) in \cite{Popkov:2005, Carrasco:2015sxh}. Moreover, the generalization to $q$-qubit Dicke states ($d=2\,, q \ne 1$) was done in \cite{Li:2015}. We extend the entanglement entropy
computation here to the general case of $q$-qudit Dicke states, with $q>0$; in fact, for the case of complex $\mathbf{q}$ where the states are given by \eqref{Dickesum2}, the entropy has a similar form as the case $q=|\mathbf{q}|$.

In order to compute the Von Neumann bipartite entanglement entropy, we begin by finding the Schmidt decomposition of the $q$-qudit Dicke states. We recall that the $d$-tuple $\vec k$ consists of non-negative integers $k_i$ satisfying $\sum_{i=0}^{d-1}k_i=n$; $\vec k$ is then referred to as a weak $d$-composition of $n$ \cite{Stanley2011}, so that we can associate all weak compositions with $q$-qudit Dicke states. For example, $\vec k=(1,2,1)$ is a weak $3$-composition of $4$. We wish to partition the state with $n$ qudits into two parts, with $l$ and $n-l$ qudits, for a positive integer $l<n$.
Thus, given $\vec k$ and $l$, we consider $d$-tuples $\vec a$ that are weak $d$-compositions of $l$ satisfying $a_i\leq k_i$. We let $\mathbb{A}^l(\vec k)$ denote the set of all such $\vec a$; explicitly,
\begin{equation}
\mathbb{A}^l(\vec k)=\left\{\,(a_0,a_1,\dots,a_{d-1})\,\,\Big|\,
0\leq a_i\leq k_i,\,\sum_{i=0}^{d-1} a_i=l\right\} \,.
\label{Alk}
\end{equation}
For example, for $\vec k=(1,2,1)$ and $l=2$, we have $\mathbb{A}^l(\vec k)=\{(1,1,0),(1,0,1),(0,1,1),(0,2,0)\}$.
If $\vec a\in\mathbb{A}^l(\vec k)$ then $\vec k -\vec a\in\mathbb{A}^{n-l}(\vec k)$, i.e. $\vec k -\vec a$ is a weak $d$-composition of $n-l$ satisfying $k_i-a_i\leq k_i$; it is now clear that the cardinality of $\mathbb{A}^l(\vec k)$ is symmetric under $l\to n-l$. With these conditions, we have essentially `cut' $\vec k$ into $\vec a$ and $\vec k-\vec a$. We can then associate the $q$-qudit Dicke states $|D_q^l(\vec a)\rangle$ and $|D_q^{n-l}(\vec k-\vec a)\rangle$ with $\vec a$ and $\vec k-\vec a$. 

The Schmidt decomposition of the $q$-qudit Dicke states for fixed $l$ can be written as 
\begin{equation}
|D_q^n(\vec k)\rangle
=\sum_{\vec a \in \mathbb{A}^l(\vec k)}
\sqrt{\lambda_q^l(\vec k,\vec a)}\,
|D_q^l(\vec a)\rangle\otimes
|D_q^{n-l}(\vec k-\vec a)\rangle \,,
\label{schmidt}
\end{equation}
with 
\begin{equation}
\lambda_q^l(\vec k,\vec a)=q^{\sum_{0\leq i<j\leq d-1}(a_i k_j-a_j k_i)}\frac{{l \brack \vec a} {n-l \brack \vec k-\vec a}}{{n \brack \vec k}} \,.
\label{lambda}
\end{equation}
A proof of this decomposition is given in Appendix \ref{sec:schmidtproof}. We note that $\sqrt{\lambda_q^l(\vec k,\vec a)}$ can be understood
as Clebsch-Gordon coefficients relating symmetric representations of $su_q(d)$.
In the limit as $q\to1,\,\lambda_q^l(\vec k,\vec a)$ reduces to the multivariate hypergeometric distribution \cite{Carrasco:2015sxh}, which has a clear combinatorial meaning: it represents the ratio of the number of permutations $w$ of $M(\vec k)$ that contain in the first $l$ positions the letters $i$ with multiplicity $a_i$ and in the last $n-l$ positions the letters $i$ with multiplicity $k_i-a_i$ to the number of all permutations $w$. 

Let $\rho_q(\vec k)=|D_q^n(\vec k)\rangle\langle D_q^n(\vec k)|$ be the density matrix of a pure $q$-qudit Dicke state, which in light of the Schmidt decomposition \eqref{schmidt} can be expressed as
\begin{equation}
\rho_q(\vec k)=
\sum_{\vec a \in \mathbb{A}^l(\vec k)}
\sum_{\vec b \in \mathbb{A}^l(\vec k)}
\sqrt{\lambda_q^l(\vec k,\vec a)\,\lambda_q^l(\vec k,\vec b)}\,
|D_q^l(\vec a)\rangle
\langle D_q^l(\vec b)|\otimes
|D_q^{n-l}(\vec k-\vec a)\rangle
\langle D_q^{n-l}(\vec k-\vec b)|\,.
\end{equation}
The reduced density matrix $\rho_q^l(\vec k)$ is obtained by tracing over the last $n-l$ qudits, which, using the orthonormality of the $q$-qudit Dicke states, forces $\vec k -\vec a=\vec k -\vec b$. So
\begin{equation}
\rho_q^l(\vec k)=\text{tr}_{n-l}(\rho_q(\vec k))=\sum_{\vec a \in \mathbb{A}^l(\vec k)}
\lambda_q^l(\vec k,\vec a)\,
|D_q^l(\vec a)\rangle
\langle D_q^l(\vec a)|\,,
\end{equation}
implying $\rho_q^l(\vec k)$ is diagonal in the $q$-qudit Dicke state basis, with eigenvalues $\lambda_q^l(\vec k,\vec a)$. The eigenvalues of a density matrix are non-negative and sum to unity, implying 
\begin{equation}
\sum_{\vec a \in \mathbb{A}^l(\vec k)}
q^{\sum_{0\leq i<j\leq d-1}(a_i k_j-a_j k_i)}
{l \brack \vec a} {n-l \brack \vec k-\vec a}={n \brack \vec k} \,,
\label{qVandermonde}
\end{equation}
which we recognize as the $q$-Vandermonde identity for multinomials.\footnote{
See e.g. Proposition 2.6 of \cite{Avalos:2020}. To convert between different $q$-deformations, see Footnote \ref{inv}.}
The Von Neumann bipartite entanglement entropy is given by
\begin{equation}
S_q^l(\vec k)=-\text{tr}(\rho_q^l(\vec k) 
\log_d \rho_q^l(\vec k))
=-\sum_{\vec a \in \mathbb{A}^l(\vec k)} \lambda_q^l(\vec k,\vec a) \log_d \lambda_q^l(\vec k,\vec a)\,,
\label{EE}
\end{equation}
where $\lambda_q^l(\vec k,\vec a)$ is given by \eqref{lambda}. For the general case of complex $\mathbf{q}$, where the states are given by \eqref{Dickesum2}, the eigenvalues of $\rho_\mathbf{q}^l(\vec k)$ are given by $|\lambda_\mathbf{q}^l(\vec k,\vec a)|$, which
simplifies to $\lambda_q^l(\vec k,\vec a)$ with
$q=|\mathbf{q}|$, see \eqref{lambda}. It follows that the entanglement entropy for complex $\mathbf{q}$ is given by \eqref{EE} with $q=|\mathbf{q}|$.

\begin{figure}[htb]
	\centering
	\begin{subfigure}{0.49\textwidth}
      \centering
\includegraphics[width=0.9\linewidth]{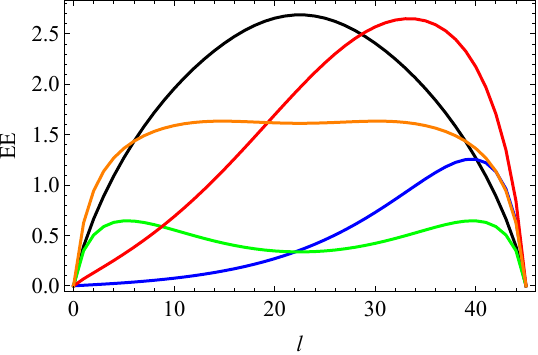}
\caption{$q=1.07$, $d=3$ and $n=45$, with\\
$\vec k_{{\rm black}}=(15,15,15)$,
$\vec k_{{\rm blue}}=(43,1,1)$,\\
$\vec k_{{\rm green}}=(1,43,1)$,
$\vec k_{{\rm red}}=(31,7,7)$,\\
$\vec k_{{\rm orange}}=(7,31,7)$.}
\label{fig:d3}
    \end{subfigure}%
    \begin{subfigure}{0.49\textwidth}
        \centering
\includegraphics[width=0.9\linewidth]{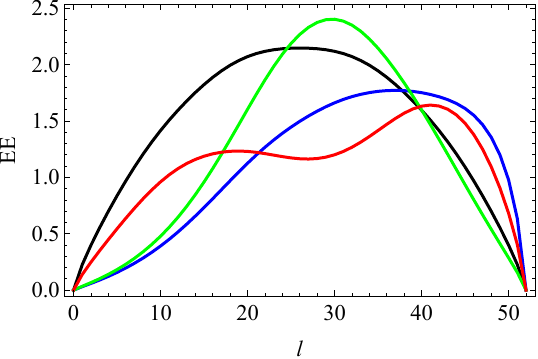}
\caption{$q=1.1$, $d=4$ and $n=52$, with\\
$\vec k_{{\rm black}}=(13,13,13,13)$,
$\vec k_{{\rm blue}}=(26,16,9,1)$,\\
$\vec k_{{\rm green}}=(26,1,9,16)$,
$\vec k_{{\rm red}}=(16,26,1,9)$
.\\}
\label{fig:d4}
	 \end{subfigure}	
\caption{Entanglement entropy (EE) \eqref{EE} of $q$-qudit Dicke states as a function of $l$.}\label{fig:EE}
\end{figure}

We compute the entanglement entropy (EE) \eqref{EE} numerically, see e.g. Fig. \ref{fig:EE}. The symmetry  $\lambda_q^l(\vec k,\vec a)=\lambda_{1/q}^{n-l}(\vec k,\vec k-\vec a)$, which follows from the $q\to1/q$ symmetry of the $q$-multinomial, induces the symmetry $S_q^l(\vec k)=S_{1/q}^{n-l}(\vec k)$ \cite{Li:2015}. We thus restrict our attention to $q\geq1$, as $q\to1/q$ simply reflects the plot over $l=n/2$. Further, we note the symmetry $\lambda_q^l(\vec k,\vec a)=\lambda_{1/q}^l({\rm rev}(\vec k),{\rm rev}(\vec a))$, corresponding to the duality symmetry \eqref{Dickesymmetry}. This implies that $S_q^l(\vec k)=S_{1/q}^l({\rm rev}(\vec k))=S_q^{n-l}({\rm rev}(\vec k))$. The latter symmetry is exemplified by the black, green and orange curves in Fig. \ref{fig:d3},
whose $\vec k$'s satisfy $\vec k={\rm rev}(\vec k)$, and are therefore symmetric about $l=n/2$. 

We also see from Fig. \ref{fig:d3} (blue vs red) that states whose $\vec k$'s have similar proportions (in this example, $k_0 \ge k_1 \ge k_2$) have EE curves (EE vs. $l$)
with similar shape. On the contrary, we see from Fig. \ref{fig:d4} that states whose $\vec k$'s have different proportions (such as blue $(26,16,9,1)$ and red $(16,26,1,9)$,  which are related by a permutation) have EE curves with different shapes. Clearly, for a given value of $n$, larger values of $d$ allow for more ways of choosing $\vec k$ with different proportions.
By suitably choosing $d$, $\vec k$ and $q$, EE curves with a great variety of shapes can be obtained.

\section{State preparation}\label{sec:algorthim}

The problem of preparing a general quantum state on a quantum computer \cite{Nielsen:2010, Mermin:2007} (i.e., by acting with unitary transformations on a simple reference state) is interesting but difficult. For the case of qubit Dicke states ($d=2\,, q=1$), an efficient deterministic algorithm has been developed, see \cite{Bartschi2019, Mukherjee:2020, Aktar:2021, Bartschi:2022} and references therein. A generalization to the case of qudits (general $d$\,, $q=1$) was recently given in \cite{Nepomechie:2023lge}. We consider here briefly the problem of preparing $\mathbf{q}$-qudit Dicke states, and find that adding a $\mathbf{q}$-dependence adds little difficulty to the algorithm. We begin by laying out the general approach for arbitrary $d$, and then work out  in detail a suitable circuit for the case of qubits ($d=2$). Finally, we show that this circuit can be simplified by removing certain gates.

\subsection{General $d$}

Similarly to \cite{Bartschi2019,Nepomechie:2023lge},  we begin by looking for a unitary operator $U_{n}$ (independent of $\vec{k}$) that generates $|D^n_\mathbf{q}(\vec k)\rangle$ for all $\vec{k}$
by acting on the identity permutation $|\re(\vec{k})\rangle$\,,
\begin{equation}
    U_n\, |\re(\vec{k})\rangle=|D_\mathbf{q}^n(\vec k)\rangle\,
    \label{qDickeOp}
\end{equation}
for all $\vec k$, 
where $\mathbf{q} = q e^{i \alpha}$ with $q=|\mathbf{q}|>0$ and $\alpha$ real.
The recursive nature of the $\mathbf{q}$-qudit Dicke states \eqref{Dickerecursiongen}
\begin{equation}
|D_\mathbf{q}^n(\vec k)\rangle
=\sum_{s=0}^{d-1}
\sqrt{\frac{[k_s]}{[n]}}
\mathbf{q}^{\frac{1}{2}\left(\sum_{r=0}^{s-1}k_r-\sum_{r=s+1}^{d-1}k_r \right)}|D_\mathbf{q}^{n-1}(\vec k-\hat s)\rangle\otimes|s\rangle \,
\label{Dickerecursiongen2}
\end{equation}
indicates that $U_n$ can be constructed recursively, as the $\mathbf{q}$-qudit Dicke states on the LHS and RHS of \eqref{Dickerecursiongen2} can be constructed by applying $U_n$ and $U_{n-1}$ to the states $|\re(\vec k)\rangle$ and $|\re(\vec k-\hat s)\rangle$ (for all $s$), respectively. This motivates finding an operator $W_n$ (independent of $\vec{k}$) that performs the mapping
\begin{equation}
W_n\,|\re(\vec k)\rangle=
\sum_{s=0}^{d-1}
\sqrt{\frac{[k_s]}{[n]}}
\mathbf{q}^{\frac{1}{2}\left(\sum_{r=0}^{s-1}k_r-\sum_{r=s+1}^{d-1}k_r \right)}|\re(\vec k-\hat s)\rangle\otimes|s\rangle \,
\label{Wsum}
\end{equation}
for all $\vec k$. In terms of this operator $W_n$, we can clearly see the recursion
\begin{equation}
U_n= \left( U_{n-1} \otimes \id \right)W_n\,.
\label{step1d}
\end{equation}
Using the initial condition $U_1=\id$, we can telescope the recursion \eqref{step1d} into a product of $W_m$ operators 
\begin{equation}
U_n =  \overset{\curvearrowright}{\prod_{m=2}^{n}} 
\left(W_m \otimes  \id^{\otimes(n-m)} \right)\,,
\label{Uresult}
\end{equation}
where the product goes from left to right with increasing $m$.
The problem therefore reduces to 
constructing quantum circuits for the $W_{m}$ operators.

\subsection{$d=2$}

\begin{figure}[htb]
	\centering
	\begin{subfigure}{0.5\textwidth}
      \centering
\begin{adjustbox}{width=0.5\textwidth, raise=7em}
\begin{quantikz}
\lstick{$0$} & \ctrl{1}  &  \gate{u(m,l)} \vqw{1} & 
\ctrl{1}  & \qw \\
\lstick{$l=1$} & \targ{} &  \ctrl{-1} & 
\targ{}   & \qw \\
\vdots &&&&\\
\lstick{$m-1$}&\qw&\qw&\qw&\qw
\end{quantikz}
\end{adjustbox}
\caption{$\RN{1}_{m,l}$ with $l=1$}
\label{fig:I1}
    \end{subfigure}%
    \begin{subfigure}{0.5\textwidth}
        \centering
\begin{adjustbox}{width=0.5\textwidth}
\begin{quantikz}
\lstick{$0$} & \ctrl{1}  &  \gate{u(m,l)} \vqw{1} & \ctrl{1}  & \qw \\
\lstick{$1$} & \qw \vqw{1} &  \qw \vqw{1} & \qw  \vqw{1} & \qw \\
\vdots &&&&\\
\lstick{$l-1$} & \qw \vqw{-1} & \ctrl{-1} \qw & \qw \vqw{-1} & \qw \\
\lstick{$l$} & \targ{} \vqw{-1} &  \ctrl{-1}  & \targ{} \vqw{-1}   & \qw \\
\vdots &&&&\\
\lstick{$m-1$}&\qw&\qw&\qw&\qw
\end{quantikz}
\end{adjustbox}
\caption{$\RN{1}_{m,l}$ with $l>1$}
\label{fig:Il}
	 \end{subfigure}	
\caption{Circuit diagrams for $\RN{1}_{m,l}$, with $u(m,l)$ defined in Eqs. \eqref{ugate}, \eqref{angles}}
\label{fig:Iops}
\end{figure}
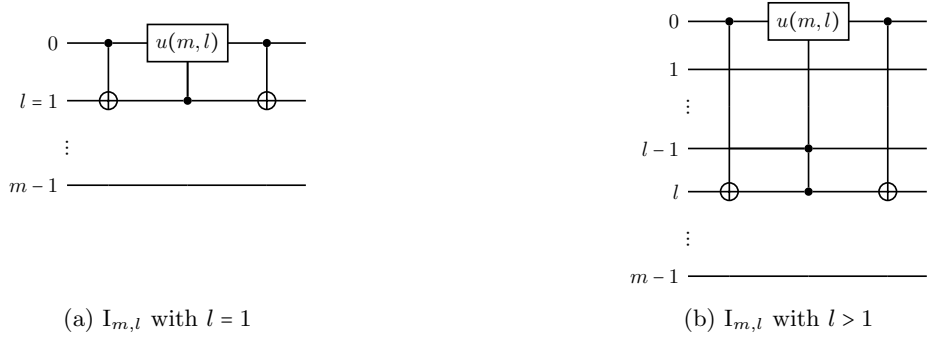

As an example, we consider the simplest case, namely $d=2$
(qubits). We associate $\vec{k}=(k_0,k_1)$ with $(n-l,l)$, so that \eqref{Wsum} reduces to finding a gate decomposition for $W_m$ ($m\leq n$) that satisfies
\begin{equation}
W_m\, |0\rangle^{\otimes (m-l)}
|1\rangle^{\otimes l}
= 
\sqrt{\frac{[m-l]}{[m]}}\, 
\mathbf{q}^{-l/2} 
|0\rangle^{\otimes (m-l-1)}
|1\rangle^{\otimes l}\otimes|0\rangle
+\sqrt{\frac{[l]}{[m]}}\, 
\mathbf{q}^{(m-l)/2} 
|0\rangle^{\otimes 
(m-l)}|1\rangle^{\otimes (l-1)}\otimes
|1\rangle  
\,,
\label{Wsum2}
\end{equation}
for all $l=1,2,\dots,m-1$ ($W_m$ acts as the identity when $l=0$ or $l=m$), where as before the brackets are given by \eqref{bracket} with $q = |\mathbf{q}| > 0$. We introduce the operator $\RN{1}_{m,l}$ acting on the $l$th, $(l-1)$th, and $0$th qubit, that performs the transformation\footnote{For $l=1$, the middle qubits in \eqref{Iaction} are omitted.}
\begin{equation}
|\underset{\mathclap{\substack{\uparrow \\ l}}}{0}\rangle\, |\underset{\mathclap{\substack{\uparrow \\ l-1}}}{1}\rangle\, 
|\underset{\mathclap{\substack{\uparrow \\ 0}}}{1}\rangle \mapsto
\sqrt{\frac{[m-l]}{[m]}}\, 
\mathbf{q}^{-l/2} 
|\underset{\mathclap{\substack{\uparrow \\ l}}}{1}\rangle\, |\underset{\mathclap{\substack{\uparrow \\ l-1}}}{1}\rangle\, 
|\underset{\mathclap{\substack{\uparrow \\ 0}}}{0}\rangle +
\sqrt{\frac{[l]}{[m]}}\, 
\mathbf{q}^{(m-l)/2}
|\underset{\mathclap{\substack{\uparrow \\ l}}}{0}\rangle\, |\underset{\mathclap{\substack{\uparrow \\ l-1}}}{1}\rangle\, 
|\underset{\mathclap{\substack{\uparrow \\ 0}}}{1}\rangle\,,
\label{Iaction}
\end{equation}
and otherwise acts as identity (as long as the $0$th qubit is in the state $|1\rangle$).
The corresponding circuit diagram is given by Fig. \ref{fig:Iops}, with one-qubit unitary $u$-gates
\begin{equation}
 u(m,l) 
 = \begin{pmatrix}
       \cos(\frac{\theta}{2}) & -e^{i \lambda}\sin(\frac{\theta}{2}) \\
       e^{i \phi}\sin(\frac{\theta}{2}) & e^{i (\phi+\lambda)} \cos(\frac{\theta}{2})
\end{pmatrix} 
\label{ugate}
\end{equation}
whose angles $\theta, \phi, \lambda$
depend on $m, l, q, \alpha$ as follows:
\begin{equation}
\theta = 2 \arccos\left(\sqrt{\frac{[l]}{[m]}} q^{(m-l)/2} \right)\,, \qquad 
\phi = \frac{m \alpha}{2} - \pi\,,  \qquad 
\lambda = -\frac{l \alpha}{2} + \pi\,.
\label{angles}
\end{equation}
In this Section, following Qiskit conventions, we label $m$-qubit vector spaces from $0$ to $m-1$, going from right to left; and in circuit diagrams, the $m$ vector spaces are represented by corresponding wires labeled from the top $(0)$ to the bottom $(m-1)$.

Note that the identity permutation 
$|\re(m-l, l)\rangle$ remains invariant under $I_{m,l'}$, i.e. 
$I_{m,l'}\,|\re(m-l, l)\rangle = |\re(m-l, l)\rangle$, if $l' \ne l$; further $I_{m,l}\,|\re(m-l, l)\rangle$ remains invariant as well,
\begin{equation}
    I_{m,l'}\, \left(I_{m,l}\,|\re(m-l, l)\rangle \right) = I_{m,l}\,|\re(m-l, l)\rangle \qquad 
    \text{ if } \qquad l' > l \,.
\end{equation}
 Hence, a quantum circuit that performs the transformation \eqref{Wsum2} for all $l=1,2,\dots,m-1$
 is given by an ordered product of such operators
\begin{equation}
W_m = 
\overset{\curvearrowleft}{\prod_{l=1}^{m-1}}
\RN{1}_{m,l} 	\,, 
\label{W2explicit}
\end{equation}
where the product goes from right to left with increasing $l$. 

\begin{figure}[htb]
	\centering
 \begin{adjustbox}{width=1.1\textwidth}
\begin{quantikz}
\lstick{$0$} \slice{} & \ctrl{1} \qw \gategroup[wires=5,steps=6,style={dashed,rounded corners,fill=blue!20, inner xsep=2pt},background]{}
&  \gate{u(5,1)} \vqw{1} & \ctrl{1} \qw & 
\ctrl{1} \qw &  \gate{u(5,2)} \vqw{1} \qw & \ctrl{1} \qw & \ctrl{1} 
\qw &  \gate{u(5,3)} \vqw{1} & \ctrl{1} \qw  & \ctrl{1} \qw 
\gategroup[wires=5,steps=3,style={dashed,rounded corners,fill=blue!20, inner xsep=2pt},background]{}
& 
\gate{u(5,4)} \vqw{1} & \ctrl{1} \qw  \slice{$W_{5}\hspace{12cm}$} & \qw \gategroup[wires=5,steps=3,style={dashed,rounded corners,fill=blue!20, inner xsep=2pt},background]{}
& \qw & \qw & \qw & \qw & 
\qw & \qw & \qw & \qw \slice{$W_{4}\hspace{9cm}$} & \qw & \qw & \qw & \qw & \qw & \qw 
\slice{$W_{3}\hspace{7cm}$} & \qw & \qw & \qw 
\slice{$W_{2}\hspace{3cm}$} & \\ 
\lstick{$1$} &\targ{}\qw &\ctrl{-1} \qw & \targ{}\qw &\qw \vqw{1} &\ctrl{-1} 
\qw & \vqw{1} \qw & \vqw{1} \qw & \qw & \vqw{1} \qw & \vqw{1} \qw & 
\vqw{1} \qw & \vqw{1} \qw & \ctrl{1} \qw & \gate{u(4,1)} & \ctrl{1} 
\qw & \ctrl{1} & \gate{u(4,2)} & \ctrl{1} & \ctrl{1} & \gate{u(4,3)} 
& \ctrl{1} & \qw & \qw & \qw & \qw & \qw & \qw & \qw & \qw & \qw &\\
\lstick{$2$} & \qw & \qw & \qw & \targ{}\qw &\ctrl{-1} 
\qw & \targ{} \qw & \vqw{1} \qw & \ctrl{-1} \qw & \vqw{1} \qw  &  
\vqw{1} \qw &  \vqw{1} \qw &  \vqw{1} \qw 
&\targ{} &  \ctrl{-1} & \targ{} \qw & \vqw{1} \qw & \ctrl{-1} \qw & 
\vqw{1} \qw & \vqw{1} \qw & \vqw{-1} \qw & \vqw{1} \qw & 
\ctrl{1} & \gate{u(3,1)} & \ctrl{1} & \ctrl{1} \qw & \gate{u(3,2)} & 
\ctrl{1} \qw & \qw & \qw & \qw &\\
\lstick{$3$} & \qw & \qw & \qw & \qw & \qw & \qw & \targ{} \qw & 
\ctrl{-1}\qw & \targ{} \qw
& \vqw{1} \qw & \ctrl{1} \qw & \vqw{1} \qw & \qw & \qw & \qw & 
\targ{} \qw & \ctrl{-1} \qw & \targ{} \qw & \vqw{1} \qw & \ctrl{-1} 
\qw & \vqw{1} \qw & \targ{} \qw & \ctrl{-1} \qw & \targ{} \qw & 
\vqw{1} \qw & \ctrl{-1} \qw & \vqw{1} \qw & \ctrl{1} \qw & 
\gate{u(2,1)} & \ctrl{1} \qw &\\
\lstick{$4$} & \qw & \qw & \qw & \qw & \qw & \qw & \qw & \qw & \qw
& \targ{} \qw & \ctrl{0} \qw & \targ{} \qw & \qw & \qw & \qw & \qw & 
\qw & \qw & \targ{} \qw & \ctrl{-1} \qw & \targ{} \qw & \qw & \qw & 
\qw &  \targ{} \qw & \ctrl{-1} \qw & \targ{} \qw & \targ{} \qw & 
\ctrl{-1} \qw & \targ{} \qw &
\end{quantikz}
\end{adjustbox}
\caption{Circuit diagram for $U_5$ \eqref{U5example}, with $W$'s (separated by red dashed lines) given by  \eqref{W2explicit} in
terms of $\RN{1}$'s in Fig. \ref{fig:Iops}. This circuit can be used to prepare the state $|D^5_{\mathbf{q}}(5-l,l) \rangle$ for any $l \in \{1, \dots, 4\}$. For the particulr case $l=3$, the shaded gates can be removed.}
\label{fig:circuitfull}
\end{figure}
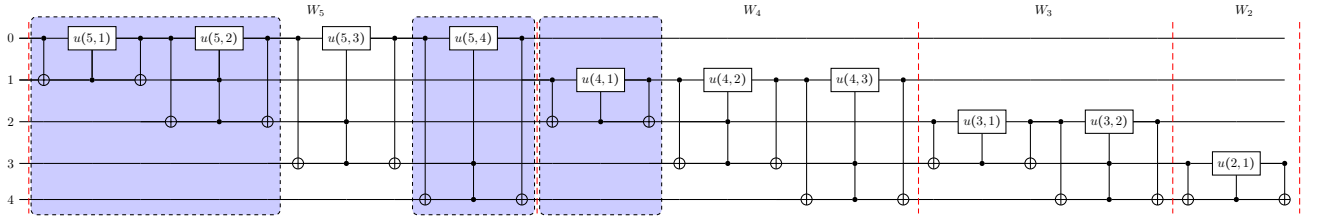

As an explicit example with $n=5$, we see from \eqref{Uresult} that
\begin{equation}
U_5 = \left(W_2 \otimes \id^{\otimes 3}\right)
\left(W_3 \otimes \id^{\otimes 2}\right)
\left(W_4 \otimes \id \right) W_5 \,,
\label{U5example}
\end{equation}
and the corresponding complete circuit diagram is shown in Fig. \ref{fig:circuitfull}. This circuit can be used to prepare the 5-qubit $\mathbf{q}$-Dicke state $|D^5_{\mathbf{q}}(5-l,l) \rangle = U_5\, |\re(5-l,l)\rangle$ from the initial state $|\re(5-l,l)\rangle$ for any $l \in \{1, \dots, 4\}$. For the particular case $l=3$, the shaded gates are redundant and
can therefore be removed, as explained below.

\subsubsection{Simplifying the operators}

While the $U_n$ operator \eqref{Uresult} in terms of  $W_m$'s \eqref{W2explicit} does generate a $\mathbf{q}$-qubit Dicke state for any $\vec{k}=(n-l,l)$, its gate count can generally be reduced. Indeed, we now proceed to prune away the redundant gates, and thereby
remain with a simplified operator $\mathcal{U}_n(n-l,l)$ in terms of corresponding simplified operators $\mathcal{W}_m(n-l,l)$, such that
\begin{equation}
    \mathcal{U}_n(n-l,l)\, |\re(n-l,l)\rangle=|D_\mathbf{q}^n(n-l,l)\rangle\,,
    \label{qDickeOpspecial}
\end{equation}
which are tailored for a fixed value of $l$. 
We begin by considering how the right-most factor in \eqref{Uresult}, $W_n$, acts on $|\re(n-l,l)\rangle$ for a fixed $l$. It is clear that we can remove $n-2$ factors in the product \eqref{W2explicit}, simplifying to $\mathcal{W}_n(n-l,l)=\RN{1}_{n,l}$. For example,
for $l=3$ in Fig. \ref{fig:circuitfull}, we can remove gates $\RN{1}_{5,1}\,, \RN{1}_{5,2}\,, \RN{1}_{5,4}$ in $W_5$.

We next consider how $W_{n-1}\otimes\id$ acts on $\RN{1}_{n,l}\,|\re(n-l,l)\rangle$. Rewriting \eqref{W2explicit} as 
\begin{equation}
W_{n-1} = 
\overset{\curvearrowleft}{\prod_{j=l+1}^{n-2}}
\RN{1}_{n-1,j}\,\, 
\overset{\curvearrowleft}{\prod_{j=l-1}^{l}}
\RN{1}_{n-1,j}\,\, 
\overset{\curvearrowleft}{\prod_{j=1}^{l-2}}
\RN{1}_{n-1,j} \,,
\end{equation}
we find that all the terms in the right-most product can be removed, 
as their controls are in qubit positions between and including $1$ 
and $l-1$, where the qubits take the value of $|1\rangle$. For 
example,  for $l=3$ in Fig. \ref{fig:circuitfull}, the term 
$\RN{1}_{4,1}$ in $W_4$ can be removed, as the controls lie on wires $1$ and $2$. Similarly, the terms in the left-most product can be removed as they also never get activated. Thus, the factor $W_{n-1}$ in \eqref{Uresult} can be simplified to
$\mathcal{W}_{n-1}(n-l,l)=\RN{1}_{n-1,l}\,\,\RN{1}_{n-1,l-1}$. 

Similar analysis can be done on the general $W_m$ factors in the product \eqref{Uresult}, leading us
to generalize \cite{Bartschi2019} to $\mathbf{q}$-qubit Dicke states by creating the quantum circuit for fixed $l$ 
\begin{equation}
\mathcal{U}_n(n-l,l) =  \overset{\curvearrowright}{\prod_{m=2}^{n}} 
\left(\mathcal{W}_m(n-l,l) \otimes  \id^{\otimes(n-m)} \right)\,,
\label{Uresultrelaxed}
\end{equation}
where
\begin{equation}
\mathcal{W}_m(n-l,l)=
\overset{\curvearrowleft}{\prod\limits_{j={\rm max}(l+m-n,1)}^{{\rm min}(l,m-1)}}
\,\RN{1}_{m,j} \,,
\label{W2explicit2}
\end{equation}
performs the transformation \eqref{Wsum2} for a fixed $l$. Note that the presence of max/min in \eqref{W2explicit2} is simply enforcing the requirement $1\leq j\leq m-1$ in $\RN{1}_{m,j}$. 

The number of $\RN{1}$-gates in $\mathcal{U}_n(n-l,l)$ is given by 
\begin{equation}
    I_n(l) = \sum_{m=2}^n \left[1+
    {\rm min}(l,m-1) - {\rm max}(l+m-n,1) \right] \,,
\end{equation}
which satisfies $I_n(l)=I_n(n-l)$, and $I_n(l) \sim l n$ for $l \ll n$. We see that there is no gate count advantage to using the dual symmetry \eqref{Dickesymmetry} $|D_\mathbf{q}^n(n-l,l)\rangle=X^{\otimes n}|D_{1/\mathbf{q}}^n(l,n-l)\rangle$ as the gate count is symmetric under $l\to n-l$. However, by using the dual symmetry to restrict to $l\le n/2$, one can minimize the separation of the wires on which the $\RN{1}$-gates act.

\begin{figure}[htb]
	\centering
 \begin{adjustbox}{width=1.0\textwidth}
\begin{quantikz}
\lstick{$0$ \quad \ket{1}}  \slice{}  & \ctrl{1}\qw &  \gate{u(5,3)} \vqw{1} & \ctrl{1} \qw  
\slice{$\mathcal{W}_{5}(2,3)\hspace{3.5cm}$} 
& \qw & \qw & 
\qw & \qw & \qw & \qw \slice{$\mathcal{W}_{4}(2,3)\hspace{7cm}$} & \qw & \qw & \qw & \qw & \qw & \qw 
\slice{$\mathcal{W}_{3}(2,3)\hspace{7cm}$} & \qw & \qw & \qw 
\slice{$\mathcal{W}_{2}(2,3)\hspace{3.5cm}$} & 
\rstick[wires=5]{$|D^{5}_{\mathbf{q}}(2,3)\rangle$}\\ 
\lstick{$1$ \quad \ket{1}} & \vqw{1} \qw & \qw & \vqw{1} \qw 
& \ctrl{1} & \gate{u(4,2)} & \ctrl{1} & \ctrl{1} & \gate{u(4,3)} 
& \ctrl{1} & \qw & \qw & \qw & \qw & \qw & \qw & \qw & \qw & \qw &\\
\lstick{$2$ \quad \ket{1}} & \vqw{1} \qw & \ctrl{-1} \qw & \vqw{1} \qw  
& \vqw{1} \qw & \ctrl{-1} \qw & 
\vqw{1} \qw & \vqw{1} \qw & \vqw{-1} \qw & \vqw{1} \qw & 
\ctrl{1} & \gate{u(3,1)} & \ctrl{1} & \ctrl{1} \qw & \gate{u(3,2)} & 
\ctrl{1} \qw & \qw & \qw & \qw &\\
\lstick{$3$ \quad \ket{0}} & \targ{} \qw & \ctrl{-1}\qw & \targ{} \qw
& \targ{} \qw & \ctrl{-1} \qw & \targ{} \qw & \vqw{1} \qw & \ctrl{-1} 
\qw & \vqw{1} \qw & \targ{} \qw & \ctrl{-1} \qw & \targ{} \qw & 
\vqw{1} \qw & \ctrl{-1} \qw & \vqw{1} \qw & \ctrl{1} \qw & 
\gate{u(2,1)} & \ctrl{1} \qw &\\
\lstick{$4$ \quad \ket{0}} & \qw & \qw & \qw
& \qw & 
\qw & \qw & \targ{} \qw & \ctrl{-1} \qw & \targ{} \qw & \qw & \qw & 
\qw &  \targ{} \qw & \ctrl{-1} \qw & \targ{} \qw & \targ{} \qw & 
\ctrl{-1} \qw & \targ{} \qw &
\end{quantikz}
 \end{adjustbox}
\caption{Circuit diagram for preparing the state
$|D^5_{\mathbf{q}}(2,3) \rangle = \mathcal{U}_5(2,3)\, |\re(2,3)\rangle$, where $\mathcal{U}_5(2,3)$ is given by \eqref{U53example} and the $\mathcal{W}$'s (separated by red dashed lines) are given by \eqref{W2explicit2} in terms of $\RN{1}$'s in Fig. \ref{fig:Iops}.}
\label{fig:circuitreduced}
\end{figure}
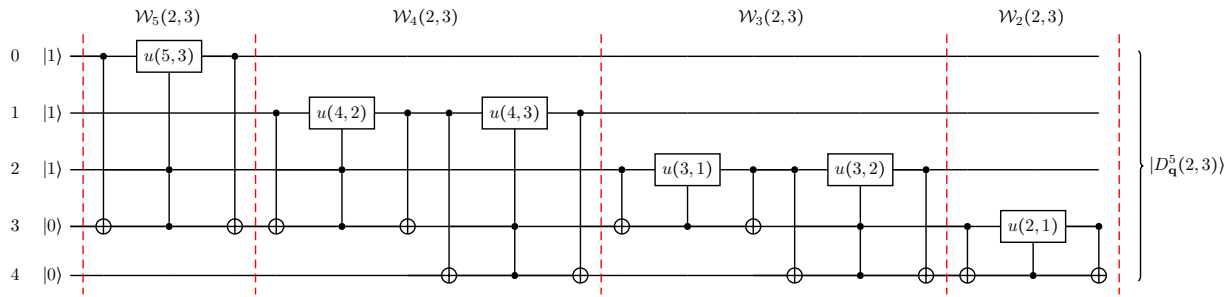

Considering as an example the particular case $(n-l,l)=(2,3)$, we see from Eqs. \eqref{qDickeOpspecial} and \eqref{Uresultrelaxed}
that $|D^5_{\mathbf{q}}(2,3) \rangle = \mathcal{U}_5(2,3)\, |\re(2,3)\rangle$, with
\begin{equation}
\mathcal{U}_5(2,3) = 
\left(\mathcal{W}_2(2,3) \otimes \id^{\otimes 3}\right)
\left(\mathcal{W}_3(2,3) \otimes \id^{\otimes 2}\right)
\left(\mathcal{W}_4(2,3) \otimes \id \right) 
\mathcal{W}_5(2,3) \,,
\label{U53example}
\end{equation}
where the $\mathcal{W}$'s are given by \eqref{W2explicit2}, as shown in Fig. \ref{fig:circuitreduced}. Note that this circuit can be obtained by removing the shaded gates from the circuit in Fig. \ref{fig:circuitfull}. An implementation in Qiskit of this and additional examples, as well as Mathematica code for verifying the results, are available as Supplementary Material.

Note that the only dependence on $\mathbf{q}$ is through the angles in the $u(m,l)$ gates \eqref{ugate}; hence, the gate count for general $\mathbf{q}$ is the same as for $\mathbf{q}=1$ \cite{Bartschi2019, Mukherjee:2020, Aktar:2021, Bartschi:2022}.
For $d>2$, we expect that the $\mathbf{q}$-dependence can be implemented in a similar way.

\section{Conclusion}\label{sec:conclusion}

We have used Quantum Groups to define the notion of $q$-qudit Dicke states through the operator formula \eqref{DickeV1}, and we have shown that these states can be expressed by the compact and useful sum formula \eqref{Dickesum}. A key ingredient of the latter is the inversion number, which is an important permutation statistic in Combinatorics. Basic properties of these states and their bipartitions are expressed by celebrated $q$-combinatorial identities \eqref{qinvidentity}, \eqref{qVandermonde}. This interplay among Quantum Groups, Combinatorics and Quantum Information suggests that there may be deeper relations among these subjects.

We have seen that the recursive nature of the $q$-qudit Dicke states \eqref{Dickerecursion} leads to a simple algorithm for their construction on a quantum computer. We note that the Schmidt decomposition \eqref{schmidt} can be regarded as a generalization of this recursion; indeed, \eqref{Dickerecursion} can be obtained by letting $l=n-1$ in \eqref{schmidt}. It should be possible to further generalize the Schmidt decomposition of $q$-qudit Dicke states to multipartite decompositions, and study multipartite entanglement. It would be interesting to see if other algorithms can be obtained by similarly exploiting these generalizations of the recursion \eqref{Dickerecursion}. It may also be possible to compute the $n \to \infty$ limit of the bipartite entanglement entropy with $q>0$, as has been done for the undeformed ($q=1$) case in \cite{Popkov:2004, Latorre:2004qn, Popkov:2005, Carrasco:2015sxh}. 

As noted in the Introduction, Dicke states have been exploited for numerous tasks in quantum information and computation. We expect that $q$-qudit Dicke states will find similar applications, but with the advantage of having available a free parameter $q$ as an additional degree of freedom. Larger values of $d$ will allow for more choices of $\vec{k}$, and therefore richer structure, see e.g. Fig. \ref{fig:EE}. As discussed in Section \ref{sec:algorthim}, the preparation of such states is not much more difficult than for undeformed qudit Dicke states.

\section*{Acknowledgements} 
 We thank Galib Hoq for his collaboration at an early stage of this 
 project, and Michelle Wachs for valuable correspondence. RN was 
 supported in part by the National Science Foundation under Grant 
 No.  PHY 2310594 and by a Cooper fellowship.

\appendix

\section{Proof of the recursion \eqref{Dickerecursion}}\label{sec:recursionproof}

We show here that the $q$-qudit Dicke state \eqref{DickeV1} satisfies the recursion \eqref{Dickerecursion}. We begin by introducing in Sec. \ref{sec:proofbasics} new operators, in terms of which the expression for the 
$q$-qudit Dicke state takes the simpler form \eqref{DickeV2}. We then evaluate this expression  in Sec. \ref{sec:mainproof} to arrive at the desired result \eqref{Dickerecursion}.

\subsection{New operators and their properties}\label{sec:proofbasics}

Let us recall that the 
single-qudit operators $H^{(1)}_i$ \eqref{Cartan1} and
$X^{\pm (1)}_i$ \eqref{Chevalley1} 
perform the mappings \eqref{Hmap} and
\eqref{Xmap}, respectively, 
on the computational basis states.
We now define new operators on $n$ qudits\footnote{We write $\mathbb{X}^{(n)}_i$ in place of $\mathbb{X}^{-(n)}_i$ for notational convenience.} 
\begin{align}
\mathbb{X}^{(n)}_i &=
X^{-(n)}_{i} X^{-(n)}_{i-1}\dots X^{-(n)}_1 \,, 
\label{Xbbrec}\\
\mathbb{H}^{(n)}_i &=\sum_{j=1}^i H^{(n)}_j \,,
\end{align}
which, for $n=1$, perform the mappings
\begin{align}
\mathbb{X}_i^{(1)}|j\rangle &=
\delta_{0j}|i\rangle \,,
\label{Xbbmap}  \\
q^{\mathbb{\pm H}^{(1)}_i/2}|j\rangle &=
\begin{cases}
q^{\pm 1/2}|j\rangle, & j=0\\ 
q^{\mp 1/2}|j\rangle, & j=i\\ 
|j\rangle, & \text{otherwise}\\ 
\end{cases} \,.
\label{Hbbmap}
\end{align}
It is natural to wonder if the $\mathbb{X}^{(n)}_i$ operators satisfy relations analogous to \eqref{Chevalleyn} and \eqref{gensrecursive}. In fact, 
\begin{align}
\mathbb{X}_i^{(n)} &=\left[\sum_{j=0}^{n-1}
\underset{\substack{\uparrow \\ 0}}{q^{\mathbb{H}^{(1)}_i/2}}
\otimes\cdots\otimes \underset{\substack{\uparrow \\ j}}{\mathbb{X}_i^{(1)}}
\otimes\cdots \otimes 
\underset{\substack{\uparrow \\ n-1}}{q^{-\mathbb{H}^{(1)}_i/2}}\right]
+ \text{rem}\,,
\label{Xbbsum0} \\
 &=
\left(q^{\mathbb{H}^{(1)}_i/2}\right)^{\otimes (n-1)}\otimes \mathbb{X}_i^{(1)}+
\mathbb{X}_i^{(n-1)}\otimes q^{-\mathbb{H}^{(1)}_i/2} + \text{rem}\,,
\label{Xbbsum} 
\end{align}
where ``rem'' denotes additional remainder terms that  vanish when applied to the state $\Sigma|0,i,i+1,\dots,d-1\rangle$, where $\Sigma|0,i,j,\dots\rangle$ denotes an arbitrary sum of kets of
$n$ qudits with each ket consisting of $0$'s, $i$'s, $j$'s, etc. 
This may be understood by substituting \eqref{Chevalleyn} into \eqref{Xbbrec}:
\begin{equation}
\mathbb{X}_i^{(n)}=\prod_{k=1}^i\left[
\sum_{j=0}^{n-1} 
\underset{\mathclap{\substack{\uparrow \\ 0}}}{q^{H^{(1)}_k/2}}
\otimes\dots\otimes
\underset{\mathclap{\substack{\uparrow \\ j}}}{X^{-(1)}_k}
\otimes\dots\otimes 
\underset{\mathclap{\substack{\uparrow \\ n-1}}}{q^{-H^{(1)}_k/2}}
\right] \,.
\end{equation}
When the products are expanded into a sum of terms, each such term will have exactly $i$ operators $X_k^{-(1)}$'s spread throughout the $n$ locations. The $n$ terms for which the $X_k^{-(1)}$'s occur at the same location together form the square-brackets term in \eqref{Xbbsum0}. The rest of the terms have an interrupted sequence of lowering operators at some location (e.g. the expansion of $\mathbb{X}_3^{(n)}$ includes terms which at some location equals
$X_3^{-(1)}q^{-H^{(1)}_2/2}X_1^{-(1)}$, which is missing $X_2^{-(1)}$) and thus acts trivially on the state $\Sigma|0,i,i+1,\dots,d-1\rangle$ using \eqref{Xmap}.

We conclude this subsection with two useful claims.

\begin{claim}
\begin{equation}
\left(q^{\mathbb{H}^{(1)}_i/2}\right)^{\otimes n}
\mathbb{X}_j^{(n)}=
\mathbb{X}_j^{(n)}\left(q^{\mathbb{H}^{(1)}_i/2}\right)^{\otimes n}
\begin{cases}
q^{-1}, & i=j\\
q^{-1/2}, & i\neq j\\
\end{cases}
\quad + {\rm rem}
\,,
\label{interchange}
\end{equation} 
\end{claim}
where ``rem'' denotes additional terms that vanish when applied to the state $\Sigma|0,j,j+1,\dots,d-1\rangle$.\footnote{We conjecture that the additional terms are absent, i.e. that the result \eqref{interchange} is actually a strict equality. However, we do not need the stronger result here.}

To show this, we compute the commutator 
\begin{equation}
\Big[\left(q^{\mathbb{H}^{(1)}_i/2}\right)^{\otimes n},
\mathbb{X}_j^{(n)}\Big]=
\Big(\sum_k 
\underset{\mathclap{\substack{\uparrow \\ 0}}}{q^{\mathbb{H}^{(1)}_j/2}}
\otimes\dots\otimes 
\underset{\substack{\uparrow \\ k}}
{\left[q^{\mathbb{H}^{(1)}_i/2}\,, \mathbb{X}_j^{(1)}\right]q^{-\mathbb{H}^{(1)}_i/2}}
\otimes\dots\otimes 
\underset{\mathclap{\substack{\uparrow \\ n-1}}}{q^{-\mathbb{H}^{(1)}_j/2}}
\Big)
\left(q^{\mathbb{H}^{(1)}_i/2}\right)^{\otimes n} + {\rm rem} \,,
\label{importantcom}
\end{equation}
where we used \eqref{Xbbsum0} (valid when acting on $\Sigma|0,j,j+1,\dots,d-1\rangle$) and that $\mathbb{H}_i^{(1)}$ commutes with $\mathbb{H}_j^{(1)}$. Now, using \eqref{Hbbmap}
\begin{equation}
\Big[q^{\mathbb{H}^{(1)}_i/2},\mathbb{X}_j^{(1)}\Big]=\mathbb{X}_j^{(1)}
\begin{cases}
q^{-1/2}-q^{1/2}, & i=j\\
1-q^{1/2}, & i\neq j\\
\end{cases} \,,
\end{equation}
acting trivially on states other than $|0\rangle$ due to \eqref{Xbbmap}. So the factor $q^{-\mathbb{H}^{(1)}_i/2}$ to the right of the commutator at location $k$ in \eqref{importantcom} necessarily produces the factor $q^{-1/2}$; thus
\begin{equation}
\Big[\left(q^{\mathbb{H}^{(1)}_i/2}\right)^{\otimes n},
\mathbb{X}_j^{(n)}\Big]=
\mathbb{X}_j^{(n)}
\left(q^{\mathbb{H}^{(1)}_i/2}\right)^{\otimes n}
\begin{cases}
q^{-1}-1, & i=j\\
q^{-1/2}-1, & i\neq j\\
\end{cases}
\quad + {\rm rem} \,.
\end{equation}
Using $AB=[A,B]+BA$, our claim \eqref{interchange} follows. $\blacksquare$

\begin{claim}
\begin{equation}\begin{split}
\label{DickeV2}
|D_q^n(\vec k)\rangle&=
\frac{1}{\sqrt{{n \brack \vec k}}}
\frac{1}{[k_1]!\dots[k_{d-1}]!}
{\mathbb{X}_1^{(n)}}^{k_1}{\mathbb{X}_2^{(n)}}^{k_2}
\dots
{\mathbb{X}_{d-1}^{(n)}}^{k_{d-1}}
|0\dots0\rangle_n.
\end{split}
\end{equation} 
\end{claim}

We can illustrate this result by expanding all of the commutators of
$\left[X^{-(n)}_{d-1},\dots \left[X^{-(n)}_2,X^{-(n)}_1 \right]\dots
\right]$ in the last factor of \eqref{DickeV1}.  One of the expanded
terms is the ordered product $X^{-(n)}_{d-1} X^{-(n)}_{d-2}\dots
X^{-(n)}_1 = \mathbb{X}^{(n)}_{d-1}$, which acting on $|0\dots0\rangle_n$ becomes
$\Sigma|0,d-1\rangle$.  The rest of the terms contain an out-of-order
product, e.g. $X^{-(n)}_{d-1} \dots X^{-(n)}_3 X^{-(n)}_1 X^{-(n)}_2$.
Each of these terms acts trivially on the state $|0\dots0\rangle_n$ in
light of \eqref{Xmap}.  
It is thus easily seen that the last factor in
\eqref{DickeV1} may be written as
\begin{equation}
{\left[X^{-(n)}_{d-1},\dots \left[X^{-(n)}_2,X^{-(n)}_1 \right]\dots \right]}^{k_{d-1}}
|0\dots0\rangle_n =  {\mathbb{X}^{(n)}_{d-1}}^{k_{d-1}}\, |0\dots0\rangle_n 
=\Sigma|0,d-1\rangle \,.
\end{equation}
The second-to-last factor in \eqref{DickeV1} acts similarly on the
state $\Sigma|0,d-1\rangle$, with the only non-trivial term being
${\mathbb{X}_{d-2}^{(n)}}^{k_{d-2}}$, producing the state
$\Sigma|0,d-2,d-1\rangle$.  
Continuing in this fashion justifies
\eqref{DickeV2}.  $\blacksquare$

\subsection{Evaluation of \eqref{DickeV2}}\label{sec:mainproof}

We now proceed to (partially) evaluate \eqref{DickeV2}, which will lead to the recursion \eqref{Dickerecursion}, repeated here for convenience
\begin{equation}
|D_q^n(\vec k)\rangle
=\sum_{s=0}^{d-1}
\sqrt{\frac{[k_s]}{[n]}}
q^{\frac{1}{2}\left(\sum_{r=0}^{s-1}k_r-\sum_{r=s+1}^{d-1}k_r \right)} |D_q^{n-1}(\vec k-\hat s)\rangle\otimes|s\rangle \,. 
\label{Dickerecursion2}
\end{equation}
Using \eqref{Xbbsum}, \eqref{Xbbmap} and \eqref{Hbbmap}, we see that
\begin{align}
{\mathbb{X}_{d-1}^{(n)}}^{k_{d-1}}
|0\dots0\rangle_n
&={\mathbb{X}_{d-1}^{(n)}}^{(k_{d-1}-1)}
\left[ 
\left(q^{\mathbb{H}^{(1)}_{d-1}/2}\right)^{\otimes (n-1)}\otimes \mathbb{X}_{d-1}^{(1)}+
\mathbb{X}_{d-1}^{(n-1)}\otimes q^{-\mathbb{H}^{(1)}_{d-1}/2}
\right] |0\dots0\rangle_n \\
&={\mathbb{X}_{d-1}^{(n)}}^{(k_{d-1}-1)}
\Big[q^{(n-1)/2}
|0\dots0\rangle_{n-1}\otimes|d-1\rangle
+q^{-1/2}
\left(\mathbb{X}_{d-1}^{(n-1)}|0\dots0\rangle_{n-1}\right)\otimes
|0\rangle\Big] \,,\nonumber  
\end{align}
where we have applied to $|0\dots0\rangle_n$
a single power of 
$\mathbb{X}_{d-1}^{(n)}$. Applying a second power gives us the terms
\begin{equation}
\begin{split}
&{\mathbb{X}_{d-1}^{(n)}}^{(k_{d-1}-2)}
\Bigg[q^{(n-1)/2}q^{1/2}
\left(\mathbb{X}_{d-1}^{(n-1)}|0\dots0\rangle_{n-1}\right)\otimes|d-1\rangle
\\&
+q^{-1/2}\left(
\left(q^{\mathbb{H}^{(1)}_{d-1}/2}\right)^{\otimes (n-1)}
\mathbb{X}_{d-1}^{(n-1)}
|0\dots0\rangle_{n-1}\right)\otimes|d-1\rangle
+q^{-2/2}
\left(
{\mathbb{X}_{d-1}^{(n-1)}}^2
|0\dots0\rangle_{n-1}\right)\otimes
|0\rangle\Bigg] \,.
\end{split}
\label{step2}
\end{equation}
We now use \eqref{interchange} to compute the interchange of 
$\left(q^{\mathbb{H}^{(1)}_{d-1}/2}\right)^{\otimes(n-1)}$ and 
$\mathbb{X}_{d-1}^{(n-1)}$; 
this single interchange produces the factor $q^{-1}$, and thus our terms \eqref{step2} simplify to 
\begin{equation}
\begin{split}
&{\mathbb{X}_{d-1}^{(n)}}^{(k_{d-1}-2)}
\Big[q^{(n-1)/2}
(q^{1/2}+q^{-1/2}q^{-1})
\left(\mathbb{X}_{d-1}^{(n-1)}|0\dots0\rangle_{n-1}\right)\otimes|d-1\rangle
\\&
+q^{-2/2}
\left(
{\mathbb{X}_{d-1}^{(n-1)}}^2
|0\dots0\rangle_{n-1}\right)\otimes
|0\rangle\Big] \,.
\end{split}
\end{equation}
It is clear that if we apply another power of $\mathbb{X}_{d-1}^{(n)}$, we will generate the term
$\left(q^{\mathbb{H}^{(1)}_{d-1}/2}\right)^{\otimes (n-1)}
{\mathbb{X}_{d-1}^{(n-1)}}^2$, which requires two interchanges, generating a factor of $q^{-2}$. 
After applying $l$ powers of $\mathbb{X}_{d-1}^{(n)}$, we have
\begin{equation}
{\mathbb{X}_{d-1}^{(n)}}^{(k_{d-1}-l)}\Bigg[
q^{(n-1)/2} c_l\,
\left({\mathbb{X}_{d-1}^{(n-1)}}^{(l-1)}
|0\dots0\rangle_{n-1}\right)\otimes|d-1\rangle
+q^{-l/2}
\left({\mathbb{X}_{d-1}^{(n-1)}}^{l}
|0\dots0\rangle_{n-1}\right)\otimes|0\rangle \Bigg] \,,
\label{lpowers}
\end{equation}
where the coefficient $c_l$ satisfies 
\begin{equation}
    c_{l+1} = q^{1/2} c_l + q^{-l/2} q^{-l}\,, \qquad l = 0, 1, \ldots \,, \qquad {\rm with} \quad c_0=0
\end{equation}
and therefore
\begin{equation}
    c_{l} = q^{(1-l)/2}\,[l] \,.
\end{equation}
Taking $l=k_{d-1}$ in \eqref{lpowers}, we obtain 
\begin{equation}
\begin{split}\label{usefuleq1}
{\mathbb{X}_{d-1}^{(n)}}^{k_{d-1}}
|0\dots0\rangle_n
&=q^{(n-1)/2}q^{(1-k_{d-1})/2} [k_{d-1}]
\left(
{\mathbb{X}_{d-1}^{(n-1)}}^{(k_{d-1}-1)}
|0\dots0\rangle_{n-1}\right)\otimes
|d-1\rangle
\\&+q^{-k_{d-1}/2}
\left(
{\mathbb{X}_{d-1}^{(n-1)}}^{k_{d-1}}
|0\dots0\rangle_{n-1}\right)\otimes
|0\rangle \,.
\end{split}
\end{equation}
Noting from \eqref{Xbbsum}, \eqref{Xbbmap} and \eqref{Hbbmap}
that 
\begin{equation}\begin{split}
&{\mathbb{X}_1^{(n)}}^{k_1}{\mathbb{X}_2^{(n)}}^{k_2}
\dots
{\mathbb{X}_{d-2}^{(n)}}^{k_{d-2}}
\left(
{\mathbb{X}_{d-1}^{(n-1)}}^{(k_{d-1}-1)}
|0\dots0\rangle_{n-1}\right)\otimes
|d-1\rangle
\\&=\left(
{\mathbb{X}_1^{(n-1)}}^{k_1}{\mathbb{X}_2^{(n-1)}}^{k_2}
\dots
{\mathbb{X}_{d-2}^{(n-1)}}^{k_{d-2}}
{\mathbb{X}_{d-1}^{(n-1)}}^{(k_{d-1}-1)}
|0\dots0\rangle_{n-1}\right)\otimes
|d-1\rangle \,,
\end{split}
\end{equation}
we see from \eqref{DickeV2} and \eqref{usefuleq1} that 
\begin{equation}\label{usefulstep}\begin{split}
|D_q^n(\vec k)\rangle&=
\sqrt{\frac{[k_{d-1}]}{[n]}}
q^{(n-k_{d-1})/2} |D_q^{n-1}(\vec k-\widehat{(d-1))}\rangle\otimes|d-1\rangle
\\&+
\frac{1}{\sqrt{{n \brack \vec k}}}
\frac{q^{-k_{d-1}/2}}{[k_1]!\dots[k_{d-1}]!}
{\mathbb{X}_1^{(n)}}^{k_1}\dots
{\mathbb{X}_{d-2}^{(n)}}^{k_{d-2}}
\left(
{\mathbb{X}_{d-1}^{(n-1)}}^{k_{d-1}}
|0\dots0\rangle_{n-1}\right)\otimes
|0\rangle \,,
\end{split}
\end{equation}
where the first term is the last term in the sum in the recursion \eqref{Dickerecursion2}. 

We now compute the effect of ${\mathbb{X}_{d-2}^{(n)}}^{k_{d-2}}$ in \eqref{usefulstep}. 
Applying one power of $\mathbb{X}_{d-2}^{(n)}$, we use again \eqref{Xbbsum} to obtain
\begin{align}
&{\mathbb{X}_{d-2}^{(n)}}^{k_{d-2}}
\left({\mathbb{X}_{d-1}^{(n-1)}}^{k_{d-1}}
|0\dots0\rangle_{n-1}\right)\otimes
|0\rangle \nonumber \\
&={\mathbb{X}_{d-2}^{(n)}}^{(k_{d-2}-1)} 
\left[ 
\left(q^{\mathbb{H}^{(1)}_{d-2}/2}\right)^{\otimes (n-1)}\otimes \mathbb{X}_{d-2}^{(1)}+
\mathbb{X}_{d-2}^{(n-1)}\otimes q^{-\mathbb{H}^{(1)}_{d-2}/2}
\right]\left({\mathbb{X}_{d-1}^{(n-1)}}^{k_{d-1}}
|0\dots0\rangle_{n-1}\right)\otimes
|0\rangle \nonumber \\
&= {\mathbb{X}_{d-2}^{(n)}}^{(k_{d-2}-1)}
\Bigg[\left(
\left(q^{\mathbb{H}^{(1)}_{d-2}/2}\right)^{\otimes (n-1)}
{\mathbb{X}_{d-1}^{(n-1)}}^{k_{d-1}}
|0\dots0\rangle_{n-1}\right)
\otimes|d-2\rangle \nonumber\\
&\qquad +q^{-1/2}\left(
\mathbb{X}_{d-2}^{(n-1)}
{\mathbb{X}_{d-1}^{(n-1)}}^{k_{d-1}}
|0\dots0\rangle_{n-1}\right)
\otimes|0\rangle
\Bigg] \,.
\end{align}
We now need to perform $k_{d-1}$ interchanges of $\left(q^{\mathbb{H}^{(1)}_{d-2}/2}\right)^{\otimes (n-1)}$ and $\mathbb{X}_{d-1}^{(n-1)}$, each of which produces a factor of $q^{-1/2}$, see \eqref{interchange}. Applying a second power of $\mathbb{X}_{d-2}^{(n)}$ will produce a term with the factor $q^{-k_{d-1}/2}q^{-1}$, where the $q^{-1}$ comes from the additional interchange of $\left(q^{\mathbb{H}^{(1)}_{d-2}/2}\right)^{\otimes (n-1)}$ and $\mathbb{X}_{d-2}^{(n-1)}$, etc. Thus
\begin{align}
&{\mathbb{X}_{d-2}^{(n)}}^{k_{d-2}}
\left(
{\mathbb{X}_{d-1}^{(n-1)}}^{k_{d-1}}
|0\dots0\rangle_{n-1}\right)\otimes
|0\rangle \nonumber\\
&\qquad=q^{(n-1)/2}q^{-k_{d-1}/2}
q^{(1-k_{d-2})/2} [k_{d-2}]
\left(
{\mathbb{X}_{d-2}^{(n-1)}}^{(k_{d-2}-1)}
{\mathbb{X}_{d-1}^{(n-1)}}^{k_{d-1}}
|0\dots0\rangle_{n-1}\right)\otimes
|d-2\rangle \nonumber\\
&\qquad\qquad+q^{-k_{d-2}/2}
\left(
{\mathbb{X}_{d-2}^{(n-1)}}^{k_{d-2}}
{\mathbb{X}_{d-1}^{(n-1)}}^{k_{d-1}}
|0\dots0\rangle_{n-1}\right)\otimes
|0\rangle \,,
\end{align}
cf. \eqref{usefuleq1}.
Substituting this result into \eqref{usefulstep} produces the second term in the recursion \eqref{Dickerecursion2}. It is not difficult to continue this argument to generate the rest of the terms in the recursion. $\blacksquare$

\section{Proof of the Schmidt decomposition \eqref{schmidt}}\label{sec:schmidtproof}

The proof of \eqref{schmidt} follows from a generalization of an 
argument below Eq. (A.9) in \cite{Carrasco:2015sxh} for $q=1$. Recall that the $q$-qudit Dicke states can be expressed by the sum formula \eqref{Dickesum}
\begin{equation}
|D_q^n(\vec k)\rangle=
\frac{1}{\sqrt{{n \brack \vec k}}}
\sum_{w\in \mathfrak{S}_{M(\vec k)}}
q^{J(\vec k)/2-\text{inv}(w)} |w\rangle_n \,.
\label{Dicksumagain}
\end{equation}
Every permutation $w$ can be split into $|w\rangle_n=|w_a\rangle_l\otimes|w_b\rangle_{n-l}$, where $|w_a\rangle_l$ and $|w_b\rangle_{n-l}$ are respectively permutations of $M(\vec a)$ and $M(\vec k- \vec a)$, where $\vec a\in\mathbb{A}^l(\vec k)$ \eqref{Alk}. We can then split the sum over permutations $w$ of $M(\vec k)$ into sums over all $\vec a\in\mathbb{A}^l(\vec k)$, over permutations $w_a$ of $M(\vec a)$, and permutations $w_b$ of $M(\vec k-\vec a)$; explicitly,
\begin{equation}
\sum_{w\in \mathfrak{S}_{M(\vec k)}}
|w\rangle_n=
\sum_{\vec a\in\mathbb{A}^l(\vec k)} 
\sum_{w_a\in \mathfrak{S}_{M(\vec a)}}
\sum_{w_b\in \mathfrak{S}_{M(\vec k-\vec a)}}
|w_a\rangle_l\otimes|w_b\rangle_{n-l}\,.
\label{permdecomp}
\end{equation}
The inversion numbers of $w$, $w_a$, and $w_b$ are related by 
\begin{equation}
    \text{inv}(w)=\text{inv}(w_a)+\text{inv}(w_b)+\sum_{i<j} (k_i-a_i)a_j \,,
    \label{invformula}
\end{equation}
where the term $\sum_{i<j}(k_i-a_i)a_j$ represents the minimum number of adjacent transpositions needed to go from  $|\re(\vec{k})\rangle$ to $|\re(\vec{a})\rangle\otimes|\re(\vec{k}-\vec{a})\rangle$, where $\re(\vec{k})$ denotes the identity permutation corresponding to $\vec k$, and similarly for $\re(\vec{a})$ and $\re(\vec{k}-\vec{a})$. Indeed, starting from the identity permutation 
\begin{equation}
|\re(\vec{k})\rangle =|
\underbrace{0\ldots 0}_{a_0}
\underbrace{0\ldots 0}_{k_0-a_0}\ldots 
\underbrace{(d-2) \dots (d-2)}_{a_{d-2}}
\underbrace{(d-2) \dots (d-2)}_{k_{d-2}-a_{d-2}}
\underbrace{(d-1) \dots (d-1)}_{a_{d-1}} \textcolor{red}{\dashline} \underbrace{(d-1) \dots (d-1)}_{k_{d-1} - a_{d-1}} \rangle \,, 
\end{equation}
we move $k_{d-2}-a_{d-2}$ letters $d-2$ from left to right (past the dashed vertical line) across $a_{d-1}$ letters $d-1$, thereby arriving at 
\begin{equation}
|\underbrace{0\ldots 0}_{a_0}
\underbrace{0\ldots0}_{k_0-a_0}\ldots 
\underbrace{(d-2) \dots (d-2)}_{a_{d-2}}
\underbrace{(d-1) \dots (d-1)}_{a_{d-1}} \textcolor{red}{\dashline} 
\underbrace{(d-2) \dots (d-2)}_{k_{d-2} -a_{d-2}}
\underbrace{(d-1) \dots (d-1)}_{k_{d-1} - a_{d-1}} \rangle \,, 
\end{equation}
which entails $(k_{d-2} - a_{d-2}) a_{d-1}$ adjacent transpositions.
We continue successively moving $k_i-a_i$ letters $i=d-3, d-4, \ldots, 0$ from left to right across all $a_j$ letters $j>i$, until we arrive at 
\begin{equation}
|\re(\vec{a})\rangle\otimes |\re(\vec{k}-\vec{a})\rangle =|\underbrace{0\ldots 0}_{a_0} \ldots \underbrace{(d-1) \dots (d-1)}_{a_{d-1}} \textcolor{red}{\dashline}
\underbrace{0\ldots 0}_{k_0-a_0} \ldots
\underbrace{(d-1) \dots (d-1)}_{k_{d-1} - a_{d-1}} \rangle \,,
\end{equation}
which entails a total number of adjacent transpositions given by $\sum_{i<j}(k_i-a_i)a_j$.

Substituting \eqref{permdecomp} and \eqref{invformula} into \eqref{Dicksumagain}, we obtain
\begin{align} 
|D_q^n(\vec k)\rangle&=
\frac{1}{\sqrt{{n \brack \vec k}}}
\sum_{\vec a\in\mathbb{A}^l(\vec k)}\Bigg\{
q^{J(\vec k)/2-J(\vec a)/2-J(\vec k-\vec a)/2-\sum_{i<j}(k_i-a_i)a_j} \nonumber\\
&\qquad\qquad\times
\Bigg(
\sum_{w_a\in \mathfrak{S}_{M(\vec a)}}
q^{J(\vec a)/2-\text{inv}(w_a)} |w_a\rangle_l
\Bigg)
\otimes\Bigg(
\sum_{w_b\in \mathfrak{S}_{M(\vec k-\vec a)}}
q^{J(\vec k-\vec a)/2-\text{inv}(w_b)} |w_b\rangle_{n-l}
\Bigg)\Bigg\} \nonumber\\
&=\sum_{\vec a\in\mathbb{A}^l(\vec k)}
\sqrt{\frac{{l \brack \vec a}{n-l \brack \vec k-\vec a}}{{n \brack \vec k}}}
q^{\frac{1}{2}\sum_{i<j}(a_i k_j-a_j k_i)}
|D^l_q(\vec a)\rangle\otimes
|D^{n-l}_q(\vec k-\vec a)\rangle \,,
\end{align}
where we have passed to the last line
using \eqref{Jresult}. $\blacksquare$


\providecommand{\href}[2]{#2}\begingroup\raggedright\endgroup

\end{document}